# Experimental Verification of 3D Plasmonic Cloaking in Free-Space

D Rainwater[1], A Kerkhoff[1], K Melin[2,3], J C Soric[3], G Moreno[3] and A Alù[3,*]
[1] Applied Research Laboratories, The University of Texas at Austin, Austin, TX 78758-4423, USA
[2] Department of Physics, The University of Texas at Austin, Austin, TX 78712, USA
[3] Department of Electrical and Computer Engineering, The University of Texas at Austin, Austin, TX 78712, USA
E-mail: alu@mail.utexas.edu, rain@arlut.utexas.edu, kerkhoff@arlut.utexas.edu

**Abstract.** We report the experimental verification of metamaterial cloaking for a 3D object in free space. We apply the plasmonic cloaking technique, based on scattering cancellation, to suppress microwave scattering from a finite-length dielectric cylinder. We verify that scattering suppression is obtained all around the object in the near- and far-field and for different incidence angles, validating our measurements with analytical results and full-wave simulations. Our near-field and far-field measurements confirm that realistic and robust plasmonic metamaterial cloaks may be realized for elongated 3D objects with moderate transverse cross-section at microwave frequencies.

## 1. Introduction

The use of metamaterials to achieve electromagnetic (EM) cloaking has been the focus of intense investigation in the past decade (see, e.g., Refs. [[1],[2]] for reviews of available approaches). Numerous techniques have been proposed to divert, suppress or feign EM scattering, using the exotic properties of several classes of metamaterials. Several of these approaches have theoretically proven that metamaterial layers may be very effective at drastically reducing the total scattering of a given object or region of space [3]–[13]. However, experimental realization has so far been limited: the available experimental proof-of-concept EM cloaks have been restricted to waveguides and 2D geometries, e.g., [14],[15]. Other experimental results on cloaking have been based on the "carpet cloaking" technique, which is limited to hide bumps on mirrors and reflectors [16]–[22]. The goal of cloaking – in fact its very definition – is however to make an object less detectable *at all angles* and for *arbitrary external excitation*. Here we report what we believe is the first successful experimental demonstration of a fully 3D cloak in free-space, verifying its functionality for near- and far-field excitations, and mapping the near-field distribution and far-field scattering over a wide angular and frequency spectra.

Our test object is a finite-length elongated dielectric cylinder. Here we apply the "plasmonic cloaking" technique [23], which has shown theoretical promise for robust response in the case of moderately sized objects. This technique is enabled by the anomalous scattering features of thin plasmonic layers with low or negative effective permittivity, which can yield drastic scattering cancellation via local negative polarizability [[23],[24]]. Using dielectric covers of negative or low-positive permittivity to reduce the scattering of objects has been studied in earlier works [25]- [27];

however, these works considered static excitation only. In this work, we experimentally realize and demonstrate the plasmonic cloaking technique [23], based on metamaterials with negative permittivity, to cloak a 3D dielectric cylinder with finite dimensions in free space to microwave excitation.

The literature on metamaterial cloaking of cylinders has dealt mostly with idealized 2D geometries: infinite cylinders, incident waves normal to the cylinder axis, and specific polarizations. Recent works [[28],[29]] have analyzed finite-length effects and excitation at oblique incidence. In particular, we have shown in [28] that plasmonic metamaterial shells can minimize scattering even under these conditions. In our recent theoretical studies, we have further emphasized the importance of coupling between transverse-magnetic $(\text{TM}_z)$ and transverse-electric $(\text{TE}_z)$ polarizations at oblique incidence, which plays a significant role in the cloak design and operation. Nevertheless, our numerical simulations gave a positive outlook for large scattering reduction over reasonable bandwidths and a broad range of incidence angles for cylinders with moderately large cross section and length comparable to the incident wavelength. Our analysis in [28] was limited to ideal homogeneous metamaterial layers, and focused only on the theoretical limitations and potential of this approach.

The present work is focused on the design, realization and testing of a realistic metamaterial plasmonic cloak, tailored to suppress microwave scattering off a dielectric cylinder. After realizing the sample, we have performed near-field mapping and far-field mono-static and bi-static measurements to verify the functionality of the realized plasmonic cloak for different excitations, indeed showing an overall significant reduction of microwave scattering both in the near- and far-fields, when excited by a microwave horn antenna placed in different positions around the object. We verify that the local phase fronts around the plasmonic cloak are nearly undisturbed by its presence, even right outside the cloak. Our far-field measurements show good comparison between theory and experimental results for various incidence angles, including both monostatic and bistatic far-field scattering. We believe that our results pave the way to realistic, practical applications of 3D stand-alone cloaks for radar evasion and non–invasive radio frequency (RF) probing [30].

**2. Plasmonic Cloak Design and Fabrication**

We focus our design on a circular dielectric cylinder of length $L = 18 \text{ cm}$, radius $a = 1.25 \text{ cm}$, permittivity $\varepsilon = 3\varepsilon_0$, and permeability $\mu = \mu_0$ (see figure 1), aiming at significantly reducing the total scattering of the cylinder around the design frequency $f_c = 3 \text{ GHz}$. An ideal, homogeneous, plasmonic cloak may be tailored to suppress the scattering from this object by applying the theory described in [[23]-[24]]; this may be realized as a thin cylindrical shell of thickness $(a_c - a)$ with effective permittivity $\varepsilon_c$ and permeability $\mu_c = \mu_0$. Even though plasmonic cloaks may be arbitrarily thin, in our design we selected a thickness of 30% the cylinder radius $\left(\frac{a_c}{a} = 1.3\right)$, so as to simplify its manufacturing and allow the use of commercially-available high permittivity dielectric materials. This implies that the cloaked object is a circular cylinder of length $L = 1.8\lambda_c$ and diameter $2a_c = 3.25 \text{ cm} = 0.325\lambda_c$, where $\lambda_c$ is the design wavelength. We have optimized our design to achieve maximum total scattering reduction of $\text{TM}_z$ polarization at normal incidence, as $\text{TM}_z$ is the dominant scattering polarization for dielectric cylinders of moderate cross section, and normal incidence has the highest total scattering cross section. In addition, in our design we have intentionally chosen a metamaterial shell with non-magnetic properties and a single-layer cloak to simplify its realization, so the only design parameter left is the cloak effective permittivity $\varepsilon_c$. As a practical matter, this choice limits our cloak to dielectric cylinders only, as cloaking conducting cylinders would require magnetic properties of the shell [28]. Single-layer plasmonic cloaks require a negative value of permittivity $\varepsilon_c$ to achieve robust scattering suppression. Indeed, the optimal cloak permittivity for the chosen thickness was determined with full-wave simulations as $\varepsilon_c = -13.6\varepsilon_0$, consistent with our theoretical results in [28].

In order to realize the required effective permittivity for the metamaterial cloak, we have explored various metamaterial geometries, such as wire media or parallel-plate implants [31]. The parallel-plate technology is particularly well-suited for cloaking $TM_z$ waves, as shown for normal incidence in [32]. This concept was also used to experimentally verify cloaking of a dielectric rod inside a waveguide [14], making the experiment effectively a 2D validation for scattering cancellation at normal incidence on an infinite cylinder [14]. Here, we extend this approach and apply it to cloak finite cylinders illuminated by $TM_z$ plane waves at arbitrary oblique angles in free space. The design requires length-spanning metallic strips extending radially outward from the core, embedded in a high-density host with permittivity $\varepsilon_s$; cf. figure 1. For normal incidence and $TM_z$ polarization with wavenumber $k_0$, the effective permittivity of the cloaking shell follows a Drude dispersion model [32]:

$$\frac{\varepsilon_c}{\varepsilon_0} = \varepsilon_s - \frac{N^2}{4k_0^2 a a_c} \quad . \tag{1}$$

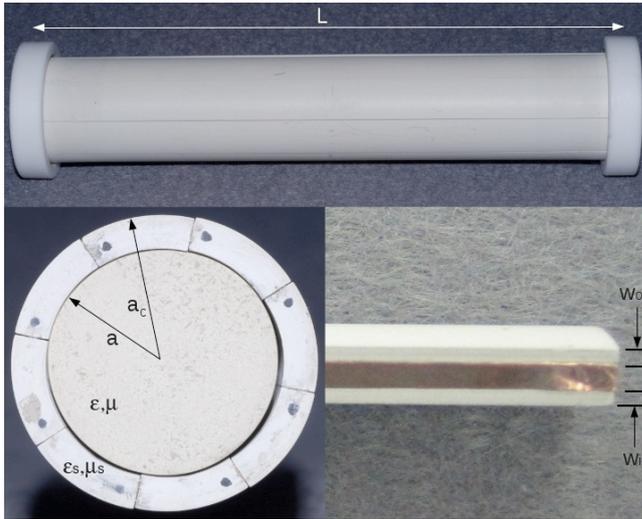

**Figure 1.** Photographs of: (top) the assembled cloak on the test cylinder with end caps; (bottom left) a cross sectional view of the assembly with end cap removed; (bottom right) a shell segment edge with copper tape used to form the metallic strip for the metamaterial cloak.

A cloak performs better in terms of isotropy in the azimuthal plane when using a larger number of strips $N$, but this comes at the price of a higher required permittivity in the host medium between neighboring implants. Critically important are the inner and outer radial gaps $w_i$ and $w_o$ between the metallic strip longitudinal edges and the cloak boundaries, as well as the strip thickness, as discussed in [32]. The strips, or fins, effectively operate as waveguide plates slightly below cutoff at the desired frequency, creating an effective low negative permittivity for the dominant mode. A finite fin thickness shortens the effective waveguide height, slightly detuning the operating frequency. This may be compensated in practice by increasing the strip edge gaps. For oblique $TM_z$ incidence, the dominant waveguide mode is unperturbed, implying that the normal component of the electric field is shorted by the metallic fins. This produces a strong spatial dispersion mechanism for $\varepsilon_c$, analogous to wire medium [33] or slit metamaterials [34]. We have fully considered these effects in our optimized design, which are rather small due to the short fin length.

We thoroughly considered these limitations, as well as the constraints associated with the practical realization of our device, optimizing the cloak design for the 3 GHz range. Using design equation (1) as a starting point, and refining the design parameters with simulations, we constructed the cloak using

8 dielectric segments with permittivity $\varepsilon_s = 16\varepsilon_0$. The copper tape used for the metallic strips has thickness 66.0 μm, separated from the inner core by an optimized gap $w_i = 0.98$ mm, and from the outer core by $w_0 = 1.3$ mm. Figure 1 shows the practical realization of the cloak components and the assembled cloaked cylinder with optimized dimensions. The test cylinder and cloak shell were both realized using machined Cuming Microwave C-Stock dielectric material. To simplify fabrication, we constructed the cloak shell as an assembly of segments (cf. figure 1) rather than as a single piece with embedded strips. The strips were precision-cut copper tape applied to one face of each segment, such that the proper edge gaps are obtained upon assembly. The segments are held together with thin end caps of low permittivity Teflon, to minimally impact the EM signature of the cylinder/cloak combination.

### 3. Full-Wave Simulations of the Realized Cloak

To predict and compare the performance of the realized cloak, we have performed an extensive set of full-wave numerical simulations for the realistic metamaterial geometry described in the previous section. Mono-static and bi-static differential scattering cross sections obtained with simulations of the realized cloak will be shown in Section 4, compared with our far-field measurements. As previously mentioned, for oblique incidence, coupling between $TE_z$ and $TM_z$ polarizations may become noticeable and cross-polarization needs to be considered in measuring the total scattering width of the object. In addition, finite length cylinders may support axial resonances when their length nears half wavelength multiples [28],[35]-[36]. Finally, our metamaterial design may be affected by strong spatial dispersion, due to the long radial implants along the cylinder axis. All these issues complicate the scattering response of the object and add interesting physics to the problem.

In order to add physical insights into the functionality of the realized cloak, and in particular how its inherent spatial dispersion properties and polarization coupling for oblique illumination may affect its overall performance, we show here the integrated results for the predicted *total* scattering of the uncloaked and cloaked objects. The total scattering cross section (SCS) curves for the cloaked and uncloaked cylinder are shown in figure 2, calculated as the integral [27]:

$$\sigma_{scat} = \int_\Omega \sigma_{scat}^d (\theta, \phi) d\Omega , \qquad (2)$$

Here, (2) is the total SCS, which includes the contributions from $TM_z$ and $TE_z$ polarizations. The differential scattering cross section, $\sigma_{scat}^d (\theta, \phi)$, i.e., the effective cross section as it appears to an observer placed in the far-field at the solid angle $(\theta, \phi)$, and the integral in (2) spans all visible angles. For monochromatic, $TM_z$ excitation the co-polarization terms are represented by the scattering coefficients $c_n^{TM}$, and the cross-polarization terms are given by the $c_n^{TE}$, as defined in [28].

It is seen that the total scattering cross section is significantly suppressed in a frequency range spanning from 3 to 3.5 GHz, with optimal performance in the 3.1–3.3 GHz range. The spatial dispersion and polarization coupling of the realized cloak have slightly shifted upwards the overall cloaking operation, still ensuring a reasonably broad bandwidth of operation and a significant total scattering reduction of up to 10 dB. As shown in the following sections, these results agree well with our near-field and far-field characterization of the realized cloak.

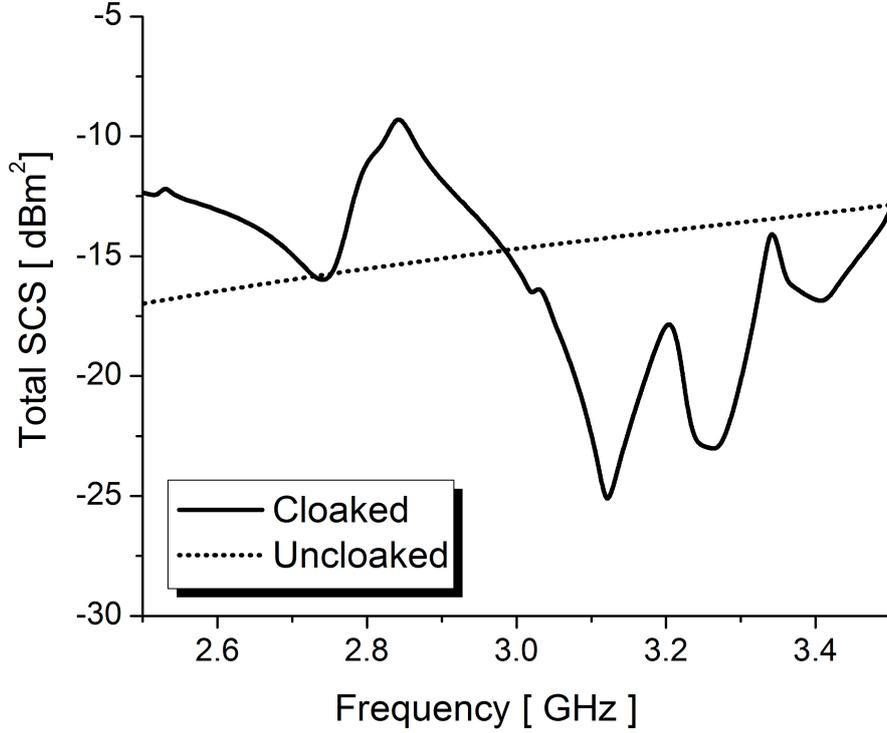

**Figure 2.** Total scattering cross section $(\sigma_{scat})$, as predicted by full-wave simulations, of the cloaked and uncloaked cylinders around the design frequency for normal incidence illumination.

## 4. Experimental Verification of 3D Plasmonic Cloaking

In order to experimentally characterize the functionality of our realized cloak, we have performed two independent sets of measurements of the cloaked object, and compared them with analogous measurements of the uncloaked object and free-space (no object). The first set of measurements was performed using a near-field microwave horn excitation, scanning and mapping with an electric-field probe the fields induced around the objects under test. The near-field measurements show the total electric field distribution along the axis of the cylinder under test, mapping spatial and temporal snapshots over a wide scan area. In a second set of measurements, we conducted far-field bi-static and mono-static scattering cross section measurements for a far-field excitation. For the mono-static measurements, the backscattering was characterized for different angles in the elevation plane. The bi-static measurements were taken for three different cases: backward and forward scattering in the elevation plane and for in- and out-of-plane azimuthal directions. The goal of this extensive campaign of measurements is to verify that the realized cloak allows for remarkable scattering reduction under different excitations and over multiple viewing angles, both in the near- and far-field regions.

*3.1 Near-field mapping for near-field excitation*
As a first experimental verification of the functionality of the realized cloak, we have performed a campaign of near-field measurements around the cloaked and uncloaked cylinders. Using an automated near-field scanning probe, we analyzed a $2\lambda_c \times 2\lambda_c$ square grid with a spatial resolution of $\Delta x = \Delta y = 0.0651 \lambda_c$. The objects under test (uncloaked and cloaked objects) were illuminated with a horn antenna producing a Gaussian beam with electric field polarized along the **z** axis, parallel to the cylinder axis placed at $2.15\lambda_c$ from the center of each object under test. Figure 3 shows the measured near-field maps of the dominant electric field component, normal to the plane of incidence, at four different frequencies, corresponding to different rows in the figure.

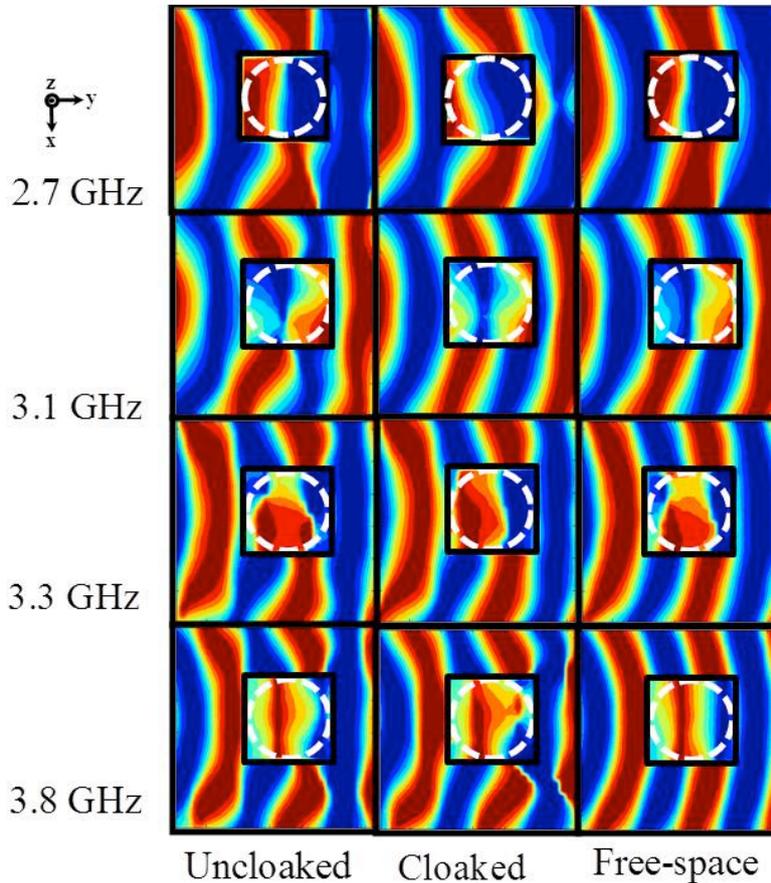

**Figure 3.** Near-field mapping of the electric field distribution (snapshot in time) around and on top of the object under test. The first and fourth rows (2.7 and 3.8 GHz, respectively) correspond to frequencies at which the cloak is expected to perform poorly. The second row (3.1 GHz) shows the lack of near-field scattering at the design frequency. The third row (3.3 GHz) corresponds to the upper-band edge of the cloak's performance.

We compare in each case the near-field distribution (snapshot in time) for the uncloaked dielectric cylinder, the cloaked cylinder and a measurement in free-space, i.e., with no object under the same experimental conditions. Each device measurement corresponds to a different column, as indicated in figure 3. The scanning was performed at a constant height, approximately 80% of the cylinder length $L$, except when the sampling plane would intersect the object under test. In this region, the probe was programmed to scan directly above the Teflon end cap of the object and is illustrated by the black squares in figure 3, which explains why in each field map a square region with discontinuous phase fronts is observed. This region allows visualizing where the object under test is placed and shows the near-field directly on top of the cylinder. The outline of the cylindrical object under test is also shown in figure 3 as the white dashed circle to demonstrate where the device or object was positioned during the near-field mapping. For consistency, the scanning setup was kept the same also in the case for which the object was not present (right column), allowing to monitor in this region of the plots the functionality of the cloak directly on top of the object.

The first and last rows of figure 3 refer to two frequencies in which the cloak is outside its range of operation (see figure 2), whereas the two central rows show the near-fields inside the cloaking frequency range. In all panels, the uncloaked cylinder (left column) significantly perturbs the Gaussian phase fronts radiated by the horn antenna in free-space (right column). However, at the two frequencies corresponding to the central rows, the cloak (middle column) is able to fully restore the

Gaussian phase fronts and suppress the large field disturbances caused by the uncloaked object. The second row shows best scattering reduction at 3.1 GHz, consistent with the simulated results in figure 2. Here, the phase fronts around and on top of the cloaked object are nearly indistinguishable from the one measured in free-space, and it is impressive to notice the large reduction in disturbance, even just outside the thin cloaking layer. Also at 3.3 GHz, at the edge of the plasmonic cloak's frequency range of operation, we find significant suppression of the near-field scattering. It is noteworthy that a significant reduction in scattering can be seen in the whole band 3.0–3.3 GHz, which demonstrates the moderately broad bandwidth of this cloaking technique and the good agreement with our modeling results given in figure 2. Outside this range, the cloaked object scatters even more than the uncloaked cylinder, as expected due to its larger physical width.

*3.2 Far-field mono-static and bi-static measurements for far-field excitation*
In our second campaign of measurements, we aimed at characterizing the cloak functionality when excitation and observation points are placed in the far-field of the device. The illuminating and receiving horn antennas were placed at a constant distance $R = 1.4 \text{ m} = 14\lambda_c$ from the object, and their position with respect to the object was rotated at different solid angles, in order to measure mono-static and bi-static differential scattering cross sections. Due to the complex channel connecting the two antennas and since the measurements were taken in a non-anechoic environment, the measurement results were post-processed in order to suppress the presence of multiple reflections using conventional software-based time gating techniques and were calibrated to the case in which no object was present.

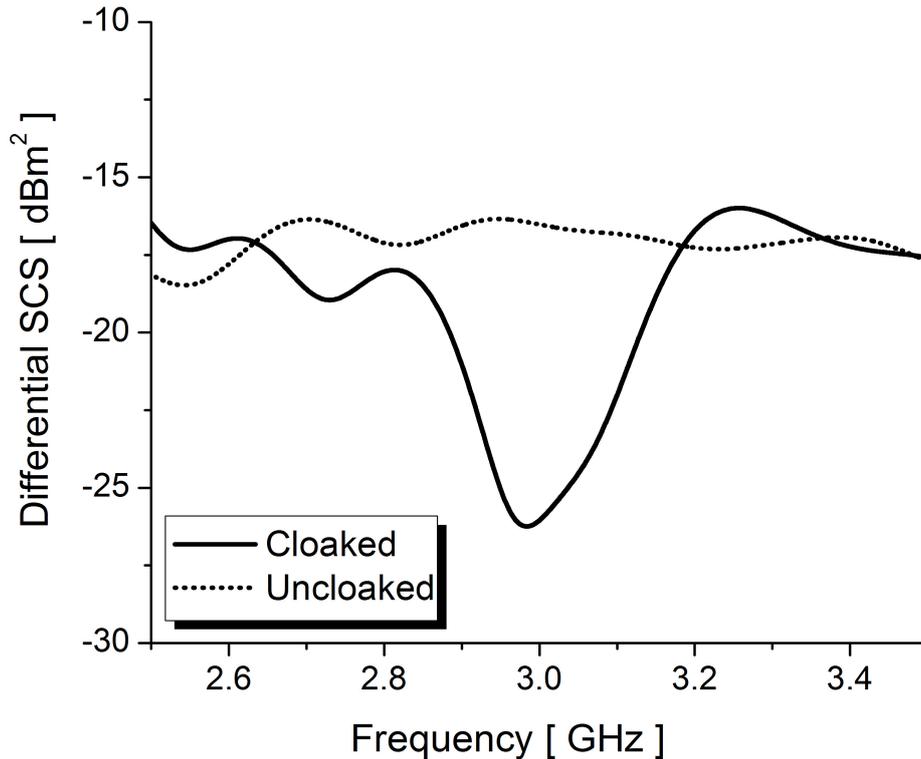

**Figure 4.** Scattering measurements for the monostatic scattering cross section of the cloaked and uncloaked object for $\sigma_{scat}^{d}(\theta = 90, \phi = 0)$, i.e., differential back scattering.

In the general case, the relative position of the transmitting and receiving antennas relative to the target is described by the two angles $\theta$ (elevation) and $\phi$ (azimuth), corresponding to the usual spherical coordinates in a reference system centered at the object position. For mono-static measurements, transmitting and receiving antennas are at the same location by definition, and the target may be simply rotated to vary the measurement angles. Since the two antennas are in the same location, and due to the symmetry of the object in the azimuthal plane, we varied only the elevation angle in these mono-static measurements.

Figure 4, as an example, shows the measured monostatic scattering cross section of the cloaked and uncloaked cylinders for $\theta = 90$ (normal incidence). The data confirms a 10 dB scattering reduction around the design frequency $f_c = 3$ GHz. We performed an extensive campaign of mono-static measurements, varying the position of the transmitting and receiving horns for the experimental background, uncloaked, and cloaked scenarios. Many mono-static oblique angle measurements were conducted, not reported here for brevity, indicating good, isotropic response on the azimuthal plane. The differences in the SCS measurements in the elevation plane $(\phi = 0, \theta)$ can be explained by the inevitable manufacturing uncertainties in fabricating the designed cloak.

We characterized the cloaking performance by normalizing the measured scattering cross section of the cloaked cylinder by the uncloaked one, which provides the 'scattering gain' figure-of-merit associated with the cloak. In parallel to these measurements, we also performed a campaign of simulations using both CST Microwave Studio and Ansoft HFSS. The two simulation codes were found to give nearly identical results; thus, only one set of curves are presented in the following corresponding to our simulations.

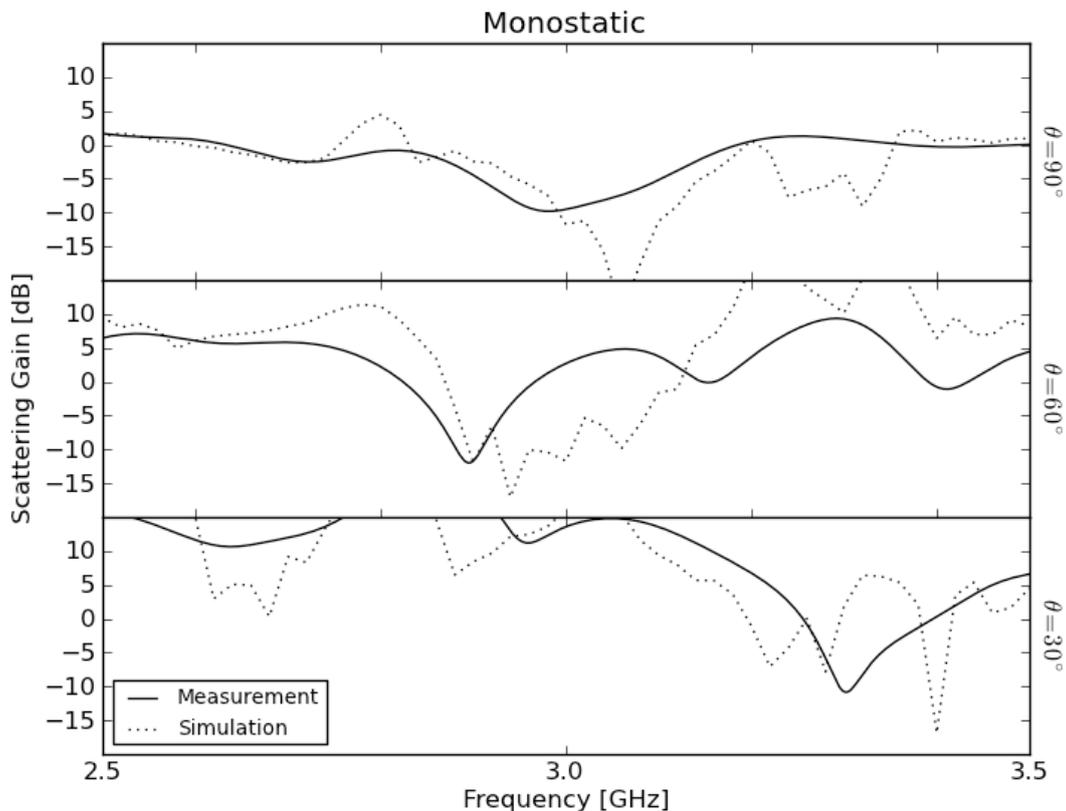

**Figure 5.** Mono-static scattering gain [dB], cloaked cylinder relative to uncloaked, for various incident elevation angles as labeled: measurement (solid) and simulated (dotted).

Figure 5 shows numerical and experimental results for the mono-static scattering gain varying the incidence angle from $\theta = 90$ (normal incidence on the cylinder) to $\theta = 30$ in steps of 30. Overall, good agreement between measurements and simulations is achieved for each illumination and over the full range of frequencies. Strong scattering suppression around the design frequency $f_c = 3$ GHz is predicted by the numerical simulations and verified by our experiment, confirming that the permittivity model (1) is accurate also for oblique incidence and that cloaking may be realistically achieved, as predicted in [28]. A maximum suppression $\geq 9.8$ dB is seen for each incidence angle. The small discrepancy in the frequency at which scattering suppression occurs as a function of incidence angle, present in both measurements and simulations, is well-predicted by our theoretical calculations and is consistent with the results in figure 2; it is associated with the TM–TE polarization coupling for incidence angles away from normal [28]. Though not shown here for conciseness, we found that the cloak provides very limited scattering suppression for grazing angles $(\theta \approx 0)$, which agrees with our predictions [28]. We stress, however, that this scenario is not so relevant for practical purposes, since the cross section and total scattering is much reduced for small values of $\theta$. It is worth stressing that our numerical simulations consider the metamaterial design of the cloak, but neglect the measurement apparatus and post-processing, which explains part of the minor differences between numerical and experimental curves. The frequency shifts at which scattering gain peaks and troughs occur between the two sets of curves may also be due to manufacturing assembly imprecision. Nonetheless, these results show that back-scattering may be strongly suppressed for several different positions of the excitation and observer. We believe that even better agreement between simulations and measurements would be obtained in an anechoic or open environment, which would remove the necessity of performing the time-gating post-processing of our measurements.

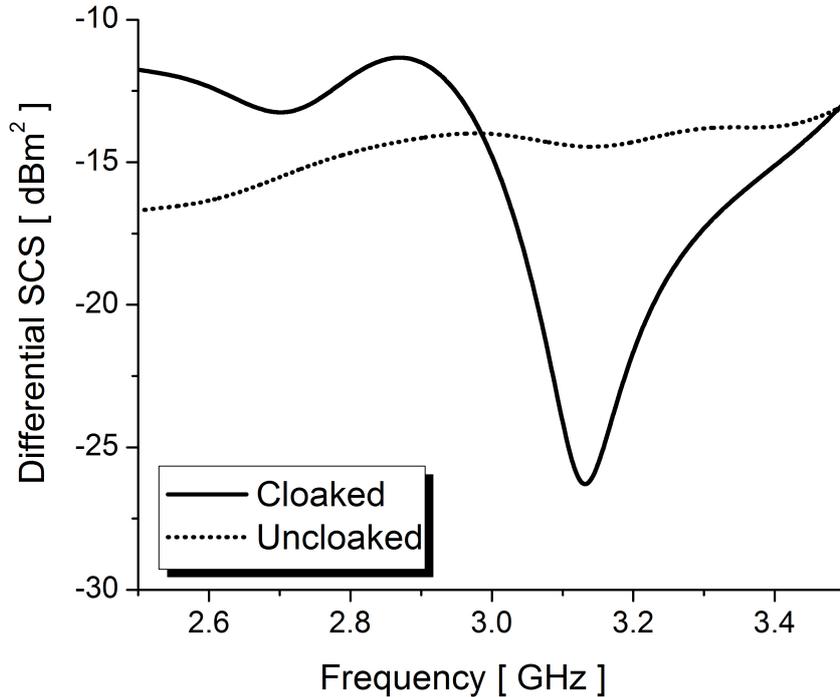

**Figure 6.** Scattering measurements for the bi-static scattering cross section for $\sigma_{scat}^d (\theta_r = -90, \phi_r = 0)$, i.e., forward differential scattering.

Bi-static results are even more interesting, since they allow us to verify that a substantial scattering reduction may be obtained for a variety of observer positions, for a given arbitrary excitation. In this second scenario, we also varied the angle $\phi_r$, measuring the azimuthal position of the receiving antenna compared to the transmitter. Figure 6, consistent with figure 4, shows the measured differential scattering cross sections of the cloaked and uncloaked cylinders for normal incidence excitation at $\theta = 90°$ and for a receiving antenna placed at $\theta_r = -90$, $\phi_r = 0$, i.e., measuring the forward differential scattering cross section. A scattering reduction $\geq 10$ dB is indeed observed around the design frequency. A collective set of bi-static results, consistent with figure 5, is shown in figures 7 and 8.

In figure 7, we show bi-static measurements scanning over the elevation angle $\theta_r$ for fixed azimuth angle $\phi_r = 0°$. Conversely, in figure 8, we varied the azimuth angle of observation for two different elevation angles, as indicated in the figure. In both cases, measurements and simulations exhibit good overall agreement. For nearly every bi-static orientation, strong suppression $\geq 7.6$ dB is observed near the design frequency. The level of discrepancy between measurements and simulations is expected to be due to manufacturing and assembly imprecision. The similarity among the curves in figure 8 for azimuthal-variation around the cylinder demonstrates a reasonably good isotropic response of the cloak, as predicted by our theoretical model [28], and the weak relevance of spatial dispersion effects due to the chosen metamaterial geometry. Variation over the polar angle (figure 7) is less isotropic, as expected, but strong scattering suppression is still observed over a wide range of angles, independent of the excitation and observation positions, again consistent with our theoretical expectations. By comparing the experimental and simulated data for the differential SCS over a large angular range, a reduction in total scattering cross section consistent with figure 2 is predicted, which confirms the functionality of our realized cloak and the realistic possibility to suppress the overall scattering of an elongated dielectric object in free-space.

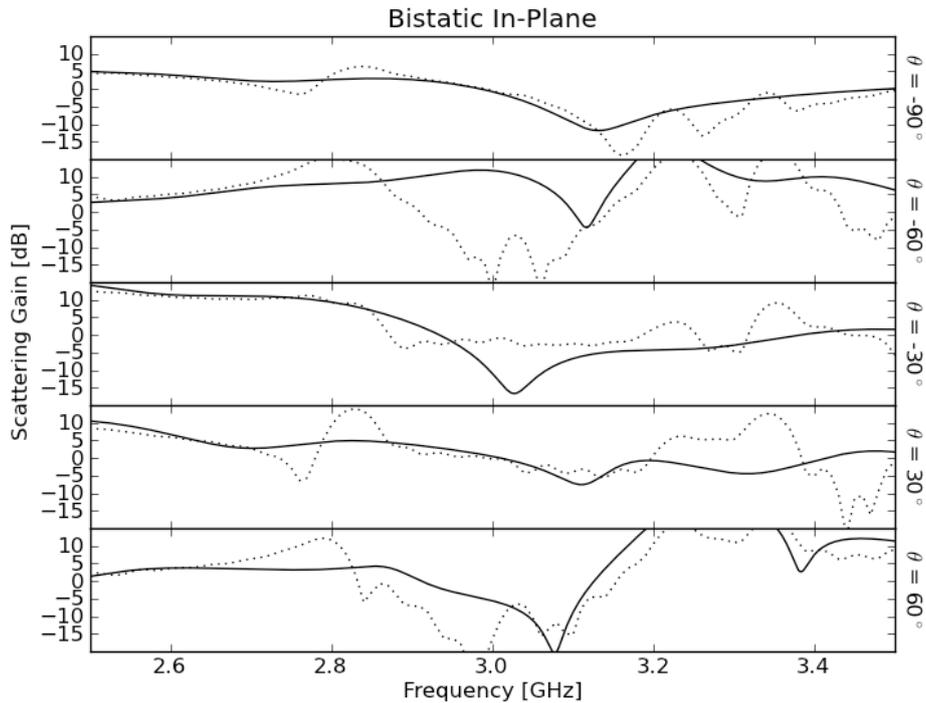

**Figure 7.** Bi-static scattering gain [dB] for normal incidence, $\phi_r = 0$ and various elevation angles: measurements (solid) and simulations (dotted) are compared in each panel.

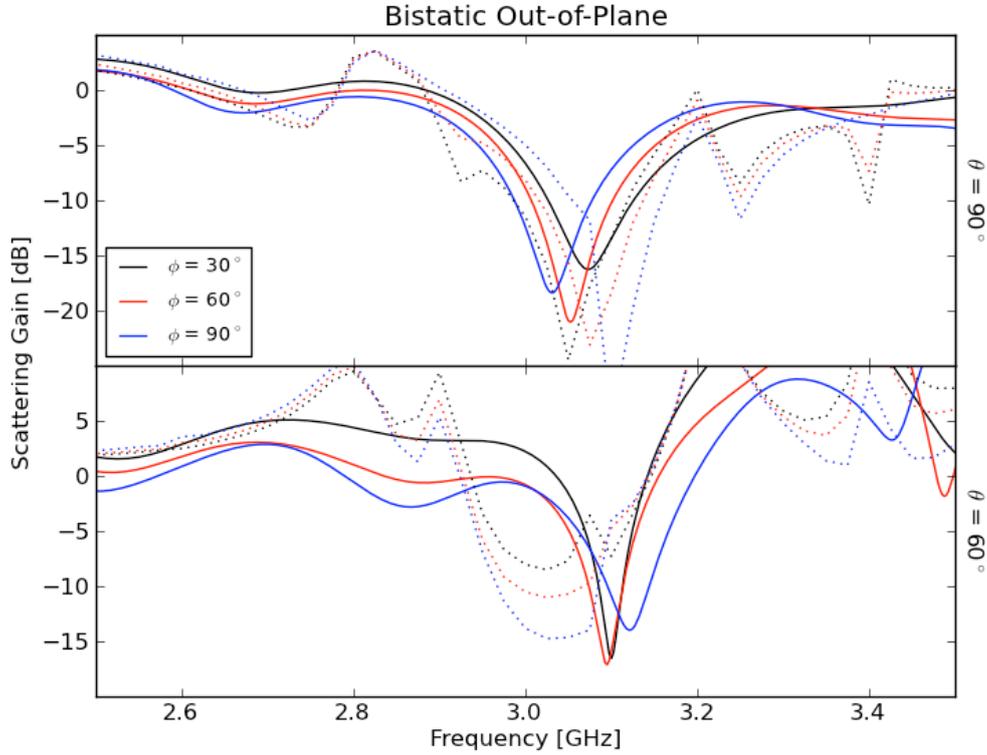

**Figure 8.** Bi-static scattering gain [dB] for several choices of incidence angle, as labeled: measurements (solid) and simulations (dotted) are compared for different in-plane and out-of-plane azimuth angles.

## 5. Conclusions

We have reported here what we believe is the first experimental verification of a 3D stand-alone cloak in free space, achieved by applying the plasmonic cloaking technique to a finite cylinder approximately two wavelengths long illuminated by microwave radiation. Our results show that robust and strong scattering suppression can be obtained over a moderately broad frequency range, weakly dependent on illumination and observation positions. Experimental measurements follow quite closely our theoretical predictions and numerical simulations [28], validating our results. Scattering is strongly reduced even for large incidence angles and near-grazing incidence for various forms of excitation, and even in the near-field. Moreover, it is evident that the extensive near-field and far-field experimental measurements are in full agreement with the total SCS as predicted from our full-wave numerical simulations of the realized plasmonic cloak reported in figure 2. The design chosen here limited the thinness of our cloak, and its functionality to one polarization. We are currently exploring alternative realizations using the mantle-cloaking technique [37]-[38], which may further reduce the overall cloak thickness and achieve scattering suppression for all polarizations.

## Acknowledgements

J.S., G.M. and A.A. were supported by the National Science Foundation (NSF) CAREER award ECCS-0953311. A. K., K. M. and D. R. were supported by an internal research award at ARL:UT. We thank Joel Banks of ARL:UT for mechanical implementation of the parallel-plate metamaterial and João Marcos from the University of Coimbra for recommendations regarding near-field mapping.